\documentstyle[sprocl]{article}
\def\journal#1#2#3#4{{#1} {\bf #2}, #3 (#4)}

\def\NPB{{\em Nucl. Phys.} B}
\def\PLB{{\em Phys. Lett.}B}

\def\be{\begin{equation}}
\def\ee{\end{equation}}
\def\bea{\begin{eqnarray}}
\def\eea{\end{eqnarray}}

\def\MPL{\em Mod. Phys. Lett.}

\def\comments#1{}

\def\half{{1\over 2}}

\def\CP{{\cal P}}

\def\a{\alpha}

\def\II{\relax{I\kern-.07em I}}

\def\IZ{\relax\ifmmode\mathchoice
{\hbox{\cmss Z\kern-.4em Z}}{\hbox{\cmss Z\kern-.4em Z}}
{\lower.9pt\hbox{\cmsss Z\kern-.4em Z}}
{\lower1.2pt\hbox{\cmsss Z\kern-.4em Z}}\else{\cmss Z\kern-.4em
Z}\fi}
\def\IB{\relax{\rm I\kern-.18em B}}

\def\ID{\relax{\rm I\kern-.18em D}}
\def\IE{\relax{\rm I\kern-.18em E}}
\def\IF{\relax{\rm I\kern-.18em F}}
\def\IG{\relax\hbox{$\inbar\kern-.3em{\rm G}$}}
\def\IGa{\relax\hbox{${\rm I}\kern-.18em\Gamma$}}
\def\IH{\relax{\rm I\kern-.18em H}}
\def\II{\relax{\rm I\kern-.18em I}}
\def\IK{\relax{\rm I\kern-.18em K}}
\def\IP{\relax{\rm I\kern-.18em P}}

\font\cmss=cmss10 \font\cmsss=cmss10 at 7pt
\def\IR{\relax{\rm I\kern-.18em R}}

\def\BR{\IR}
\def\BZ{\IZ}
\def\BR{\IR}

\def\log{{{\rm log}\,}}
\def\logx#1{{{\rm log}\,{\left({#1}\right)}}}

\def\lim{{lim}}

\input epsf

\def\SUSY#1{{{\cal N}= {#1}}}                   
\def\lbr{{\lbrack}}                             
\def\rbr{{\rbrack}}                             

\def\inv#1{{1\over{#1}}}                              

\def\MR#1{{{\BR}^{#1}}}               

\def\MR#1{{{\BR}^{#1}}}               
\def\MS#1{{{\bf S}^{#1}}}               
\def\MT#1{{{\bf T}^{#1}}}               
\def\CP#1{{{\bf P}^{#1}}}              
\def\MF#1{{{\bf F}_{#1}}}               

\def\ev#1{{\langle {#1} \rangle}}           

\def\trp#1{{{\rm tr}\{ {#1} \} }}            
\def\trr#1#2{{{\rm tr}_{#1}\{ {#2} \} }}            

\def\rep#1{{{\bf {#1}}}}                      

\def\hepth#1{{\it hep-th/{#1}}}

\def\frac#1#2{{{{#1}}\over {{#2}}}}           

\def\u{{\mu}}
\def\v{{\nu}}

\def\lam{{\lambda}}

\def\Lam{{\Lambda}}

\begin{document}
\title{Correlation functions of the global $E_8$ symmetry currents
       in the Heterotic 5-brane theory.}
\author{Morten Krogh}
\address{Department of Physics, Jadwin Hall, Princeton University, \\
Princeton, NJ 08544, USA}

%
%

\maketitle
\abstracts{We consider the 5-brane placed at one end of the 
world in the Heterotic $E_8 \times E_8$ theory. The low energy 
 theory is a 6 dimensional $(1,0)$ superconformal theory with 
$E_8$ as a global symmetry. We calculate the two-point correlator 
of the $E_8$ current in 6 dimensions and in 4 dimensions after 
compactification on $\MT{2}$. This correlator is derived in 3 
different ways: From field theory, from 11 dimensional 
supergravity and from F-theory.}

This paper is a written version of the talk given at the 
1998 Trieste Conference on
Superfivebranes and Physics in 5+1 Dimensions. The talk 
presented the work of ~\cite{CGK}, where a more elaborate 
 discussion can be found.

\section{Introduction}
In the past two years, many examples of nontrivial IR fixed
points in various dimensions have emerged.
Some of the most exciting ones are the 5+1D chiral theories.
The first of such theories with $\SUSY{(2,0)}$ SUSY has been discovered
in \cite{WitCOM} as a sector of type-IIB compactified on an $A_1$
singularity. A dual realization was found in \cite{StrOPN}
as the low-energy description of two 5-branes of M-theory.
Another theory of this kind arises as an M-theory
5-brane approaches the 9-brane \cite{GanHan},\cite{SWSIXD}.
When the distance between
the 5-brane and 9-brane is zero, the low-energy is described
by a nontrivial 5+1D fixed point. This theory is chiral
with $\SUSY{(1,0)}$ SUSY and a global $E_8$ symmetry.
In \cite{SWSIXD},\cite{WitV} more examples of $\SUSY{(1,0)}$
theories have been given.
We will use the terminology of \cite{WitV} and call the $E_8$ theory
$V_1$.
Many other 5+1D theories have been recently
constructed in \cite{BluInt},\cite{IntNEW}.

String theory is a powerful tool to study such theories.
The idea is to identify a dual description such that quantum
corrections of the original theory appear at the classical level
of the dual \cite{KV}.
The toroidal compactification of the
$\SUSY{(1,0)}$ 6D theory (and hence 4D $\SUSY{2}$ QCD) can be studied
using the brane-probe technique discovered in \cite{SeiPRB},\cite{BDS}.
The world-volume theory on a brane probe in a heterotic
string vacuum (which is quantum mechanically corrected)
is mapped by duality to a world-volume theory on a brane inside a curved
background which is not quantum mechanically corrected.
This allows one to determine the low-energy behavior in 4D.
At the origin of the moduli space
one obtains an IR fixed point with $E_8$ global symmetry.

The purpose of the present work is to extract information about
the local operators of such theories.
The $E_8$ theory $V_1$ 
has a local $E_8$ current $j^a_\u(x)$ ($a=1\dots 248$
and $\u=0\dots 5$).
We will be interested in the correlator  $\ev{j^a_\u(x) j^b_\v(0)}$.
The strategy will be to couple the theory to a weakly coupled $E_8$
gauge theory and calculate the effect of $V_1$ on the $E_8$ 
coupling constant.
We will study the question both for the 5+1D theory and for
the 3+1D conformal theories.
We will present three methods for evaluating the correlator.
The first method is purely field-theoretic and applies to the
3+1D theories. Deforming the theory with a relevant operator
one can flow to the IR where a field-theoretic
description of $SU(2)$ or $U(1)$ with several quarks \cite{GMS} can be
found.
This will allow us to determine the correlator as a function
on the moduli space.
From this function we can deduce the high-energy behavior
of the correlator and find out how many copies of the $E_8$
theory can be gauged with an $E_8$ SYM before breaking 
asymptotic freedom.

The other two methods for determining the correlators involve 
M-theory and F-theory.
The gravitational field of a 5-brane of M-theory which
is close to a 9-brane changes
the local metric on the 9-brane. After compactification
on a large $K3$ this implies that the volume of the $K3$ at the 
position of the 9-brane is affected by the distance from the 5-brane
(see \cite{WitNWT}). This can be interpreted as a dependence of the $E_8$
coupling constant on the VEV which specifies the position of the 5-brane.
From this fact we can extract the current correlator.
The third method involves the F-theory \cite{VafaFT} realization of the $E_8$
theory \cite{SWSIXD},\cite{WitPMF},\cite{MVII}.
The $V_1$ theory is obtained in F-theory compactifications on a 
3-fold by blowing up a point in the (two complex dimensional) base.
By studying the effect of the size of the blow-up on the size of
the 7-brane locus we can again determine the dependence of the 
$E_8$ coupling constant on the VEV.

The paper is organized as follows.
In section (2) we calculate the current-current correlators in 3+1D
using field theory arguments and we argue that 10 copies of the
$E_8$ theory can be coupled to a gauge field.
In section (3) we study the effect of a 5-brane on the volume of
a 9-brane in M-theory and deduce the correlator from this setting.
In section (4) we present the F-theory derivation.
In section (5) we conclude.

\section{Field theory derivation of the 4 dimensional correlator}

In this section we will derive the form of the $E_8$ current-current
correlator for the $E_8$ conformal theory and as a result
we will argue that in 4D one can couple up to 10 copies of the $E_8$
theory to a $\SUSY{2}$  $E_8$ Yang-Mills gauge theory.

We start with the $E_8$ conformal theory in 4 dimensions whose Seiberg-Witten
curve is given
by \cite{GMS}:
\be
y^2 = x^3 + u^5,
\label{swcee}
\ee
$u$ parameterizing the moduli space of the Coulomb branch.
We are looking for an expression of the form 
\be
\ev{j^a_\u (q) j^b_\v(-q)} = (q^2 \eta_{\u\v} - q_\u q_\v ) \delta^{ab}
f(q^2, u, \Lambda),\qquad a,b=1\dots 248,
\label{fqul}
\ee
where $q$ is the momentum and $\Lambda$ is some fixed UV-cutoff.
This UV-cutoff is not physical. It is just an artifact of the
 Fouri\'er transform. The space-time correlator 
$\ev{j^a_\u (x) j^b_\v(y)}$ does not require a cutoff.

To determine the form of $f$ in \ref{fqul} for $q^2 = 0$
we can couple the $E_8$ SCFT to a weakly coupled $E_8$ gauge field
and ask how the $E_8$ coupling constant changes as a function of $u$.
When the $E_8$ coupling constant is very small the coupling
 does not change the curve \ref{swcee} by much.
For a generic value of $u$ the massless modes of the $E_8$ SCFT
are neutral under the global $E_8$ and the charged matter 
has a typical energy of order $u^{1/6}$.
The $\ev{jj}$ correlator will modify the low energy $E_8$ coupling constant
to the form
$$
{1\over {g(u)^2}} = {1\over {(g_0)^2}} + f(q^2 = 0, u, \Lambda),
$$
where $g_0$ is the bare coupling constant.
On the other hand, standard renormalization arguments require
that it should be possible to re-absorb the $\Lambda$ dependence in
the bare coupling constant. Thus, dimensional analysis restricts
the form of $f(0,u,\Lambda)$ to 
\be
f(0,u,\Lambda) = c\, \logx{ {\Lambda \over {|u|^{1/6}}} }.
\label{contha}
\ee
 
To determine $c$ we deform the theory by adding a relevant
operator to its (unknown) Lagrangian such as to break the
global $E_8$ symmetry down to $D_4$ ($SO(8)$) by putting Wilson lines
on the torus.
The advantage is that the $D_4$ conformal fixed point 
can be analyzed  in standard field-theory. It is the IR free
theory of $SU(2)$ coupled to 4 massless quarks \cite{GMS}.

The global $E_8$ of the original theory has been broken by
the operators to a global $SO(8)$.
For the $SO(8)$ theory we can ask what is
\be
\ev{j_\u^A(q) j_\v^B(-q)},\qquad A,B = 1\dots 28,
\label{jjsof}
\ee
where $A,B$ are $SO(8)$ indices.
The point is now that we can calculate this correlator
for the $SO(8)$ theory from field theory. The relevant 
field theory is $SU(2)$ gauge theory with 4 quarks. 
From this correlator we can extract the original 
$E_8$ correlator. The details can be found in \cite{CGK}.
We conclude that for the $E_8$ theory
\be
\ev{j^a_{\mu}(q)  j^b_{\nu}(-q)} =
   - \frac{3 C({\rm fund.})}{4 \pi^2}   
\delta^{ab}
 (q^2 \eta_{\mu\nu} -q_{\mu}q_{\nu})\,
  \log\left|\frac{(\Lambda \lam^2)^{1/3}}
  {u^{1/6}}\right|. 
\label{ghi}
\ee
This means that the value of $c$ in \ref{contha} is 
$$
c = - 3 \frac{C({\rm fund.\ of\ SO(8)})}{4 \pi^2} 
  = -  \inv{10} \times\frac{C_2(E_8)}{4 \pi^2}.
$$
This value of $c$ implies that 10 copies of the $E_8$ SCFT
can be coupled to an $E_8$ SYM.\cite{CGK}

\section{Derivation from M-theory}

The system of the $(1,0)$ $E_8$ theory ($V_1$) coupled to $E_8$ SYM
can be realized in M-theory as a 5-brane which is close to the 9-brane.
The modes of the $V_1$ theory come from the 5-brane bulk and from
membranes stretched between the 5-brane and 9-brane while the $E_8$
SYM comes from the 9-brane  bulk. Let us compactify on $K3\times\MT{2}$.

The effect that we are trying to study corresponds to the following
question.
The gravitational field of the 5-brane affects the metric at the
position of the 9-brane. Thus, as we change the distance of the 5-brane
from the 9-brane the volume of the $K3$ changes as a function of $x$
\cite{WitNWT}.
The volume of $K3\times\MT{2}$ is related to 
the 3+1D $E_8$ coupling constant.
In field-theory, 
this is interpreted as a running of the 
$E_8$ coupling constant as a result of the change of the VEV of the 
$V_1$ theory.

We apply the general setting and formulae of \cite{WitNWT}
to the case where
the distance of the 5-brane from the 9-brane is much smaller
than the compactification scale of $K3\times\MT{2}$
and calculate the effect.

We must also mention that the after compactification of the
system of a 5-brane and 9-brane on $\MS{1}$ we get 4-branes near
8-branes. This setting has been studied in \cite{SeiFIV}, in the context
of brane probes, where a related effect is observed. 
The position of the probe affects the value of a classical field,
in that case the dilaton, which is then re-interpreted as a 1-loop
effect in field theory.
In fact, the relation between the classical
supergravity calculation and the 1-loop field-theory
calculation follows from perturbative string-theory.
The 1-loop result is a loop of DD strings connecting the 4-brane
to the 8-brane while the classical supergravity result is the same diagram
viewed from the closed string channel.

\subsection{Geometrical setup and review}

In this section we will examine the theory of a 5-brane in 
M-theory on $\MR{5,1} \times K3 \times \MS{1}/\BZ_2$ and review
some relevant facts from \cite{HWII}, \cite{WitNWT}.

The geometric setup is as follows.
The coordinates $ (x^1 , x^2, ... ,x^6) $ parameterize $\MR{5,1}$,
$(x^7 , x^8 , x^9  ,x^{10} )$
parameterize K3 and finally $x^{11}$ parameterize $\MS{1}/\BZ_2$.
All 5-branes have their world-volume along $\MR{5,1}$ and  are located
at a point
in $K3 \times \MS{1}/\BZ_2$. All configurations will be defined on the
whole ${\bf S^1}$ and are symmetric under the $\BZ_2$ (working
``upstairs'' -- in the terminology of \cite{HWII}). This means,
for example, that every time there is a 5-brane between the two fixed planes
of the $\BZ_2$ there is also a mirror 5-brane. There would be an
equivalent formulation (``downstairs'') where configurations
were only defined on the interval between the two ``ends of the world''.

We know that M-theory on $\MR{9,1} \times \MS{1}/\BZ_2$ is heterotic
$E_8 \times E_8$
with one $E_8$ theory living on each fixed plane of the $\BZ_2$.
If we compactify
this theory on K3 we need to supply a total of 24 instantons and 5-branes.
The theory we are interested in is a single
5-brane coupled to an $E_8$ gauge theory.
To achieve this we need to have no instantons in one of the $E_8$
theories and one 5-brane close to this ``end of the world''.
The remaining 23 instantons and 5-branes must therefore be either
instantons in the other ``end of the world'' or 5-branes in the bulk.

In the 6-dimensional description the distance of the 5-brane from the
``end of the world,'' $x$, is a modulus.
The effective gauge coupling of the $E_8$ depends on $x$.
 From the 6-dimensional point of view certain degrees of freedom connected
to the 5-brane act as matter coupled to the $E_8$ gauge field.
Since the couplings and masses of this matter depend on $x$,
the low energy effective $E_8$ gauge coupling, $g$, will depend on $x$.
Here we will calculate the $x$-dependence of $g$ from M-theory or more
precisely from 11-dimensional supergravity. For supergravity to be
applicable all distances involved in the problem need to be much
bigger than the 11-dimensional Planck scale. This means especially
that ${\rm Vol}(K3) \gg l_{Planck}^4$.
Furthermore we are interested in the behaviour of the theory when it is
close to the point with tensionless strings or equivalently with a zero
size instanton, which is $x = 0$.
To be in that situation we take $x \ll vol(K3)^{\inv{4}}$.
 The $x$-dependence of the 6-dimensional gauge coupling $g$,
comes about because the volume of the K3 at $X^{11}=0$ depends on $x$.

To calculate $g$ we need to find the form of the metric as a function
of $x$. The calculation of the metric and of $g$ is described 
in detail in \cite{CGK}. Here we will 
just state the result, which is
\be
\inv{g^2} =   \inv{g_0^2}  - \inv{8 \pi^{2}} x T_2.
\label{gxt}
\ee
Here $T_2$ is the tension of the membrane in 11 dimensions.
The expression $x T_2$ is the tension of the strings in the six-dimensional
theory. This is because the membrane is stretched with one direction
along the 11th direction and two directions along $\MR{5,1}$.

Compactifying further down to 4 dimensions on a torus of area $A$
is straightforward
\be
\inv{g^2} =   \inv{g_0^2}  - \inv{8 \pi^{2}} A x T_2.
\label{gax}
\ee
This equation contains the needed information about the 
4 dimensional theory. It tells how the gauge coupling in 
a $E_8$ gauge theory runs as a result of coupling to 
the 4 dimensional superconformal theory with $E_8$ global 
symmetry. In \cite{CGK} it is shown how this implies that 
10 of these saturate the $\beta$-function in complete 
agreement with the field theory derivation.

\section{The 6D current-current correlator from F-theory}

In this section, we use the duality between F theory on elliptic
Calabi-Yau 3-folds and Heterotic String on $K3$ to compute the
effective gauge coupling of heterotic string in six dimensions.
We shall see that the result agrees completely with the
corresponding  M-theory
calculation to first order.  A second order effect which is
suppressed by a factor of the volume of the K3 and by the 
length of $\MS{1}/\BZ_2$ in calculations
in the previous section naturally emerges in the F-theory setting.
In the limit in which we extract the correlator for $V_1$, i.e. taking
the volume of K3 and the size of $\MS{1}/\BZ_2$ to infinity,
this second order effect vanishes.

We start with $V_1$ and couple it to a 6D $E_8$ SYM theory.
The gauge theory is defined with a UV cut-off, but this imposes no
problem for us since all we need is the dependence of the
IR coupling constant on the VEVs of the $V_1$ theory.
To be precise, we take the $E_8$ UV cut-off to be $\Lam$
and fix the $E_8$ coupling constant at $\Lam$.
The Coulomb branch of the $V_1$ theory has a single tensor multiplet.
We denote the VEV of its scalar component by $\phi$.
$\phi$ is the tension of the BPS string in $\MR{5,1}$.
In M-theory $\phi = x T_2$.
The mass scale of the $V_1$ theory is thus $\phi^{1/2}$.
We would like to find the dependence of the IR $E_8$ coupling constant
on $\phi$ when $\phi \ll \Lambda$. Heuristically speaking,
the running $E_8$ coupling constant will receive contributions
from loops of modes from $V_1$ of mass $\sim \phi^{1/2}$.

The set-up that we have just described arises in the heterotic string
compactified on $K3$ with a small $E_8$ instanton.
We take the $(0,23)$ embedding with a single 5-brane in the bulk
close to the 9-brane with unbroken $E_8$. The F-theory dual
has a base $B$ which is the Hirzebruch surface
$\MF{11}$ with one point blown-up  \cite{SWSIXD},\cite{MVII}.
$\MF{n}$ is a $\CP{1}$ bundle over $\CP{1}$. Let the area of 
the fiber $\CP{1}$ in $\MF{11}$
(i.e. the K\"ahler class integrated over the fiber)
be $k_F$ and the area of zero section $\CP{1}$ of the fibration
be $k_D$.

We blow-up a point in the zero section of the fibration of $\CP{1}$
over $\CP{1}$.
There are 10 7-branes wrapping that zero-section and passing
through a point of the exceptional divisor. These are responsible
for the unbroken $E_8$ gauge group. Let $k_E$ be the area of the
exceptional divisor. The area of the above mentioned
7-brane locus (part with unbroken $E_8$) is $k_D$.
The K\"ahler class is
$$
k = (k_F - k_E) E + k_F D + (k_D + k_E - n k_F) F
$$
where $E,D,F$ are the cohomology classes of the exceptional divisor,
base and fiber.
\bea
E\cdot E = -1, &\qquad
E\cdot D = F\cdot D = 1,\\
D\cdot D =  n-1, &\qquad
 E\cdot F = F\cdot F = 0.\\
\label{intscns}
\eea

A 3-brane wrapping the exceptional divisor gives a BPS string
in $\MR{5,1}$ (corresponding to the membrane connecting
the M-theory 5-brane to the end of the world).  Its tension is
given by integrating the D3-brane tension over $E$.  Using \cite{Polch}
$$
2\kappa^2 \tau_p^2 =2\pi(4\pi^2 \alpha')^{3-p}
$$
the tension of the BPS string is simply
$$
\phi= \pi^{1/2} k_E
$$
in the units $\kappa=1$.
The volume of the whole base is
\be
V = \half k\cdot k = k_F (k_D + k_E) -\half k_E^2 -{n\over 2} k_F^2.
\label{volbas}
\ee
This volume is the 6D inverse gravitational constant and we have
to keep it fixed.
Although the $V_1$ modes have an effect on the gravitational constant
as well, by dimensional analysis, this effect is much smaller
than $\phi$ and behaves as $\sim \phi^2$.
How should $k_F$ depend on $\phi$, in our setting?
$k_F$ measures the tension of 3-branes wrapped on $F$.
On the heterotic side, these are elementary
strings which occupy a point on K3.
Their tension is fixed in the heterotic picture. Thus $k_F$
is independent of $\phi$.

Now we come to the gauge coupling.
To do this calculation it is convenient to imagine that $E_8$ is 
broken down to $U(8)\subset E_8$. 
The gauge kinetic term for 8 unwrapped 7-branes
of the same type is
$$
\int \tau_7 {{(2\pi \a')^2}\over {4}} \trr{\rep{8}}{F^2}\, d^8 x.
$$
We are working in the conventions
$$
\trp{T^a T^b} = \delta^{ab}, \qquad a,b=1\dots 248.
$$
For the $U(8)$ subgroup this means that
$$
\trr{\rep{8}}{T^a T^b} = {1\over {2}}\delta^{ab}.
$$
This means that for a configuration of 10 7-branes forming
an $E_8$ gauge theory the gauge kinetic term is:
$$
{1\over 8}\int (2\pi \a')^2\, \tau_7\,
\left(\sum_{a=1}^{248} F^a F^a\right)\, d^8 x.
$$
From this we read off (in units where $\kappa = 1$)
$$
{1\over {4g^2}} = {1\over 8}(2\pi \a')^2\,\tau_7
= {1\over {32}}\pi^{-3/2}.
$$
Wrapping the 7-branes on $D$ we get a 5+1D $E_8$ gauge theory with
coupling constant 
$$
{1\over {4g^2}} 
= {1\over {32}}\pi^{-3/2} k_D.
$$
 From \ref{volbas} we find that when $V$ and $k_F$ are kept fixed
and $k_E = \pi^{-1/2} \phi$,
the $E_8$ coupling constant is
\be
{1\over {g(\phi)^2}} = {1\over {8}}\pi^{-3/2}\lbr
(k_D + k_E) - k_E \rbr = {1\over {(g_0)^2}}
-{1\over {8\pi^2}} \phi.
\label{rung}
\ee
We have used the fact that $(k_D + k_E)$ is fixed to first order
in $\phi$ when $V$ is fixed.  The other two terms in $V$ are higher
order corrections dual to taking $K3$ and the distance between
the ends of the world to be large in the  M-theory calculations.
Eqn.\ref{rung} describes the running of the $E_8$ coupling constant
because  of the coupling to $V_1$.
This is in complete agreement with the result \ref{gxt} obtained from M-theory.

\section{Discussion}

We have found that for the 3+1D $E_8$ super-conformal theory with
Seiberg-Witten curve
$$
y^2 = x^3 + u^5,
$$
the 2-point $E_8$ current correlator on the Coulomb branch
satisfies:
$$
\ev{j_\u^a(q) j_\v^b(-q)} = \cases{
{{C_2(E_8)}\over {40\pi^2}} \delta^{ab} (q_\u q_\v -q^2 \eta_{\u\v})
   \logx{ {\Lambda\over |u|^{1/6}} }
& for $|q| \ll |u|^{1/6}$
\cr
{{C_2(E_8)}\over {40\pi^2}} \delta^{ab} (q_\u q_\v -q^2 \eta_{\u\v})
   \logx{ {\Lambda\over |q|} }
& for $|q| \gg |u|^{1/6}$
\cr
}
$$
where $\Lambda$ is a UV cutoff which is an artifact of Fouri\'er
transforming.

We deduced that 10 copies of the $E_8$ theory can be coupled as ``matter''
to an $\SUSY{2}$ $E_8$ SYM gauge field.

In 5+1D we found the expression
for the low-energy limit of the 5+1D correlator of
the $\SUSY{(1,0)}$ $E_8$ theory on the Coulomb branch and away from
the origin:
$$
\ev{j_\u^a(q) j_\v^b(-q)} = 
-{{C_2(E_8)}\over {240\pi^2}}\,
   \delta^{ab} (q^2 \eta_{\u\v} - q_\u q_\v)
   (\Lambda^2 - \phi)\qquad
{\rm for\ } |q| \ll \phi.
$$
where $\phi$ is the VEV of the scalar of the low-energy tensor multiplet.

It would be interesting to determine the correlator in the UV region
$|q| \gg |\phi|$ or, equivalently, at the fixed point $\phi = 0$.
It seems that the methods presented in this paper are not powerful
enough for that purpose.

\section*{Acknowledgments}
I wish to thank Yeuk-Kwan E. Cheung and Ori J. Ganor with 
whom this work was done.
I furthermore wish to thank Steve Gubser, Igor Klebanov, Sangmin Lee,
Sanjaye Ramgoolam and Savdeep Sethi for discussions.
I wish to thank the organizers of the 1998 Trieste Conference on
Superfivebranes and Physics in 5+1 Dimensions for letting 
me speak at the conference.
This work was supported by the Danish Research Academy.

\end{document}